\documentclass{egpubl}
\usepackage{eg2018}

\ConferencePaper        
 \electronicVersion 

\ifpdf \usepackage[pdftex]{graphicx} \pdfcompresslevel=9
\else \usepackage[dvips]{graphicx} \fi

\PrintedOrElectronic

\usepackage{t1enc,dfadobe}

\usepackage{egweblnk}
\usepackage{cite}
\usepackage{amssymb}
\usepackage{amsmath}

\newcommand{\pcpnet}{\textsc{PCPNet}}
\newcommand{\mypara}[1]{\paragraph*{#1.}}

\title[\pcpnet: Learning Local Shape Properties from Raw Point Clouds]%
      {\pcpnet \\ Learning Local Shape Properties from Raw Point Clouds}

\author[P. Guerrero, Y.Kleiman, M. Ovsjanikov \& N. J. Mitra]{
\parbox{\textwidth}{\centering
Paul Guerrero$^1$
\hspace{20pt}
Yanir Kleiman$^2$
\hspace{20pt}
Maks Ovsjanikov$^2$
\hspace{20pt}
Niloy J. Mitra$^1$
}\\
{\parbox{\textwidth}{\centering
$^1$University College London
\hspace{20pt}
$^2$LIX, \'{E}cole Polytechnique, CNRS
}}}

\begin{document}


\maketitle


\begin{abstract}
   In this paper, we propose \pcpnet, a deep-learning based approach for estimating local 3D shape properties in point clouds. In contrast to the majority of prior techniques that concentrate on global or mid-level attributes, e.g., for shape classification or semantic labeling, we suggest a patch-based learning method, in which a series of local patches at multiple scales around each point is encoded in a structured manner. Our approach is especially well-adapted for estimating local shape properties such as normals (both unoriented and oriented) and curvature from raw point clouds in the presence of strong noise and multi-scale features. Our main contributions include both a novel multi-scale variant of the recently proposed PointNet architecture with emphasis on local shape information, and a series of novel applications in which we demonstrate how learning from training data arising from well-structured triangle meshes, and applying the trained model to noisy point clouds can produce superior results compared to specialized state-of-the-art techniques. Finally, we demonstrate the utility of our approach in the context of shape reconstruction, by showing how it can be used to extract normal orientation information from point clouds.

\ccsdesc[500]{Computing methodologies~Point-based models}
\ccsdesc[300]{Computing methodologies~Shape analysis}
\ccsdesc[300]{Computer systems organization~Neural networks}

\printccsdesc   
\end{abstract}

\section{Introduction}
\label{sec:introduction}

A fundamental problem in shape analysis is to {\em robustly estimate local shape properties directly from raw point clouds}. Although the problem has been extensively researched, a unified method that is robust under various data imperfections (e.g., varying noise level, sampling density, level of details, missing data) remains elusive. 

In the context of continuous surfaces, local surface properties such as normals and curvature are classical differential geometric notions~\cite{carmo_76} and are known to uniquely characterize local geometry up to rigid transformations.
In the context of discrete surfaces (e.g., polygonal meshes), the estimation methods fall broadly in two groups: evaluate normal/curvatures using discrete differential geometry operators, or use local primitive fitting and `read off' normal/curvatures using attributes from the fitted primitives. 

\begin{figure}[t!]
    \centering
    \includegraphics{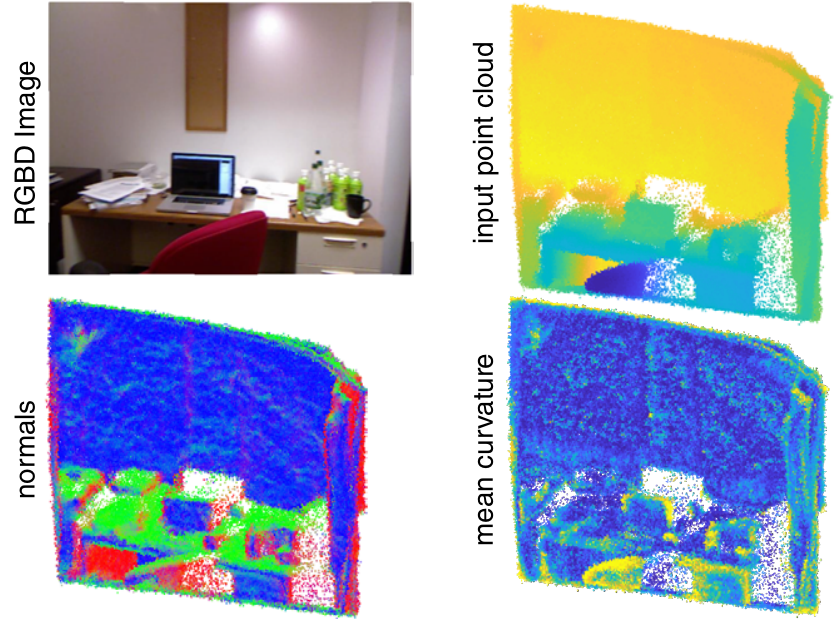}
    \caption{Our method can jointly estimate various surface properties like normals and curvature from noisy point sets. This scene is from the NYU dataset~\cite{Silberman:ECCV12} with added Gaussian noise and the bottom row shows properties estimated with \pcpnet. Note that our method was not specifically trained on any RGBD dataset.
    \label{fig:motivation}
    \vspace{-15pt}
    }
\end{figure}

In the case of point clouds, a first challenge is the lack of connectivity information. Further, in real acquisition setups (e.g., using a depth camera), the data is often noisy, incomplete, and typically exhibit varying sampling density (based on scanning direction). Figure~\ref{fig:motivation} shows a typical raw point cloud scan. The usual approach is to first locally define a neigborhood around each point using a $r$-radius ball, and then fit a local primitive (e.g., plane or quadric surface) to the neighboring points. Although the method is simple and regularly used, in practice, the challenge is to appropriately pick the various parameters. For example, while it is desirable to pick a small value of $r$ to avoid smoothing out sharp features, the same parameter can lead to unstable results in the presence of noise. Similarly, a large value of $r$ can over-smooth local features, although the estimated values tend to be more stable under noise. 
The key challenge for such fitting-based methods is to {\em manually set} the various parameters depending on the (unknown) feature distribution in shapes and (unknown) noise margins in the raw scans.

In this work, we propose a data-driven approach for estimating local shape properties directly from raw pointclouds. 
In absence of local connectivity information, one could try to voxelize the occupied space, or establish local connectivity using $k$ nearest neighbor graph. However, such discretizations introduce additional quantization errors leading to poor-quality local estimates. In an interesting recent work on normal estimation, Boulch et al.~\cite{Boulch:2016} generate a set of local approximating planes, and then propose a data-driven denoising approach in the resultant Hough-space to estimate normals. In Section~\ref{sec:results}, we demonstrate that even such a specialized approach leads to errors on raw point clouds due to additional ambiguities introduced by the choice of the representation (RANSAC-like Hough space voting).

Inspired by the recently introduced PointNet architecture~\cite{pointnet17} for shape classification and semantic segmentation, we propose a novel multi-scale architecture for robust estimation of local shape properties under a range of typical imperfections found in raw point clouds. 
Our key insight is that local shape properties can be robustly estimated by suitably accounting for shape features, noise margin, and sampling distributions. However, such a relation is complex, and difficult to manually account for. Hence, we propose a data-driven approach based on local point neighborhoods, wherein we train a neural network \pcpnet\ to directly learn local properties (normals and curvatures) using groundtruth reference results under different input perturbations. 

We evaluated \pcpnet\ on various input point clouds and compared against a range of alternate specailized approaches. Our extensive tests indicate that \pcpnet\  consistently produces superior normal/curvature estimates on raw point clouds in challenging scenarios. Specifically, the method is general (i.e., the same architecture simultaneously produces normal and curvature estimates), is robust to a range of noise margins and sampling variations, and preserves features around high-curvature regions. 
Finally, in a slightly surprising result, we demonstrate that a cascade of such local analysis (via the network depth) can learn to robustly extract {\em oriented normals}, which is a {\em global} property (similar patches may have opposite inside/outside directions in different shapes).
Code for \pcpnet\ is available at \url{geometry.cs.ucl.ac.uk/projects/2018/pcpnet/}.

\section{Related Works}
\label{sec:related_work}

\subsection{Estimating local attributes} 

\mypara{Normal estimation}
Estimating local differential information such as normals and curvature has a
very long history in  geometry processing, motivated in large part by 
its direct utility in shape reconstruction. Perhaps the simplest and best-known
method for normal estimation is based on the classical Principal Component
Analysis (PCA). This method, used as early as \cite{Hoppe:1992:SRU:142920.134011}, is based on analyzing the
variance in a patch around a point and reporting the direction of minimal
variance as the normal. It is extremely efficient,  has been analyzed extensively, and its behavior
in the presence of noise and curvature is well-understood
\cite{mng_est_normal_ijcga_04}. At the same time, it suffers from several key
limitations: first it depends heavily on the choice of the neighborhood size,
which can lead to oversmoothing for large regions or sensitivity to noise for
small ones. Second, it is not well-adapted to deal with structured noise, and
even the theoretical analysis does not apply in the case multiple, interacting,
noisy shape regions, which can lead to non-local effects. Finally, it does not
output the normal \emph{orientation} since the eigenvectors of the covariance
matrix are defined only up to sign.

Several methods have been proposed to address these limitations by both
designing more robust estimation procedures, capable of handling more
challenging data, and by proposing techniques for estimating normal
orientation. The first category includes more robust distance-weighted
approaches \cite{pauly2003shape}, methods such as osculating
jets~\cite{Cazals:2003:EDQ} based on fitting higher-order primitives robustly,
the algebraic point set surface approach based on fitting algebraic
spheres~\cite{guennebaud:PSS},  approaches based on analyzing the
distribution of Voronoi cells in the neighborhood of a point
\cite{amenta1999surface,alliez07,Merigot:2011:VCF}, 
and those based on edge-aware resampling \cite{huang2013edge} among many others. 
Many of these techniques come with strong theoretical approximation and
robustness guarantees (see e.g., Theorem 5.1 in
\cite{Merigot:2011:VCF}). However, in practice they require a careful setting of
parameters, and often depend on special treatment in the presence of strong or
structured noise. Unfortunately, there is no universal
parameter set that would work for all settings and shape types.

\mypara{Normal orientation} These challenges are arguably even more pronounced
when trying to estimate \emph{oriented} normals, since they depends on both local (for direction)
and global (for orientation) shape properties. Early methods have relied on simple greedy orientation
propagation, as done, for example in the seminal work on shape reconstruction
~\cite{Hoppe:1992:SRU:142920.134011}. However, these approaches can easily fail in
the presence of noisy normal estimates or high complexity shapes. As a result, they have 
has been extended significantly in recent years both through more robust
estimates \cite{huang2009consolidation}, which have also been adapted to handle large missing regions through  point skeleton estimation \cite{wu2015deep}, and global techniques based on signed
distance computations \cite{mullen2010signing}, among others. Nevertheless,
reliably estimating oriented normals remains challenging especially across
different noise levels and shape structures.

\begin{figure*}[t!]
	\includegraphics[width=1.0\textwidth]{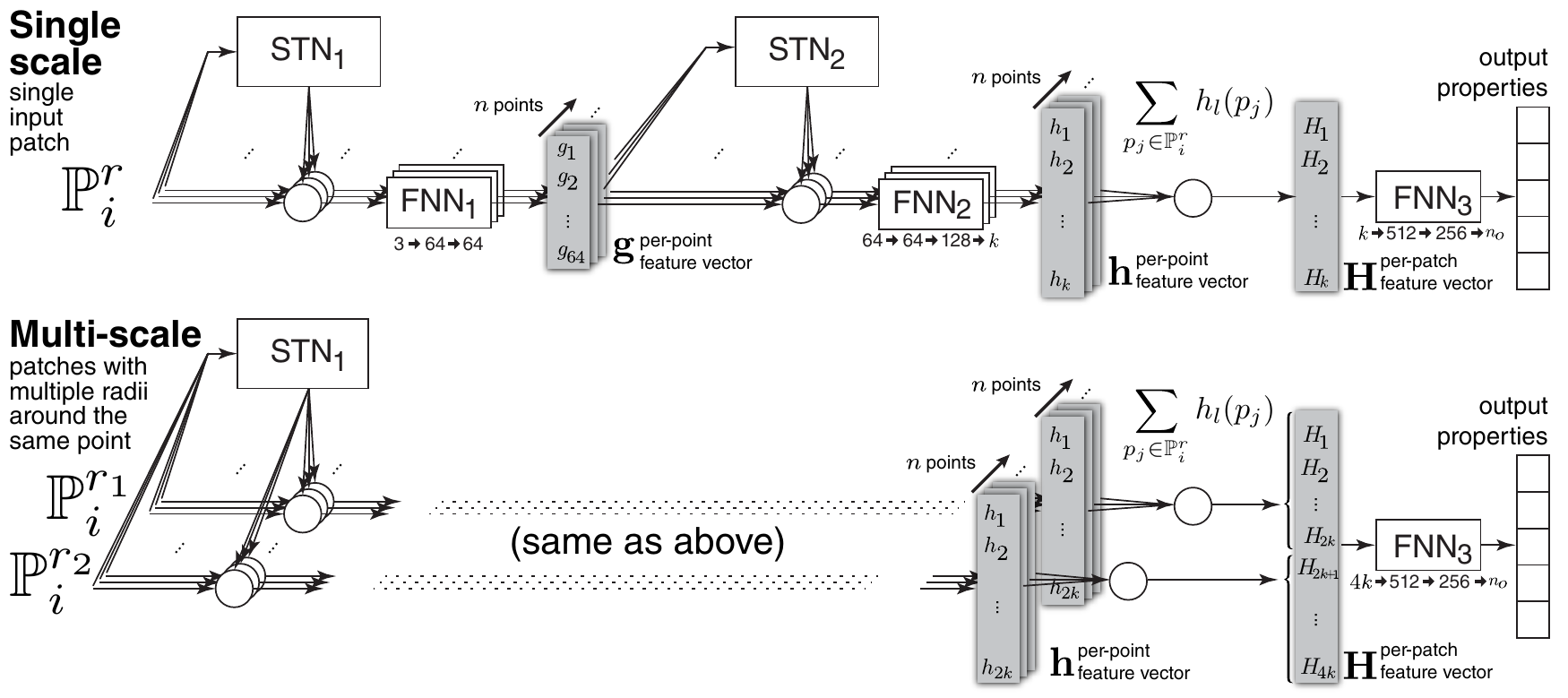}
	\caption{\footnotesize We propose two architectures for our network, a single- and a multi-scale architecture. In both architectures, the network learns a set of point functions $h$ in the local space of a point set patch. Similar to a density estimator, point function values at each point in the patch are summed up. The resulting per-patch feature vector $H$ can be used to regress patch properties. STNs are spatial transformer networks that apply either a quaternion rotation ($\textrm{STN}_1$) or a full linear transformation ($\textrm{STN}_2$). FNNs are fully connected networks. In the multi-scale architecture patches of different scale are treated as a single larger patch by the point functions. The final sum is performed separately for each patch.}
	\label{fig:arch}
\end{figure*}

\mypara{Curvature estimation} Similarly to surface normals, a large number
of approaches has also been proposed for estimating principal
curvatures. Several such techniques rely on estimating the normals first and
then robustly fitting \emph{oriented} curvatures
\cite{huang2002combinatorial,lange2005anisotropic}, which in turn lead to
estimates of principal curvature values. In a similar spirit, curvatures can be
computed by considering normal \emph{variation} in a local neighborhood
\cite{berkmann1994computation,kalogerakis2009extracting}. While accurate in the
presence of clean data, these techniques rely on surface normals, and any error
is only be exacerbated by the further processing. One
possible way to estimate the normal and the second fundamental form jointly is by directly fitting higher order polynomial as is done for example in
the local jet fitting approach \cite{Cazals:2003:EDQ}. This approach is fast and
reliable for well-structured point sets, but also requires setting a scale
parameter which is challenging to find for all types of noise or shape types. In Section~\ref{sec:results}, we compare our \pcpnet\  to jet fitting in a wide variety of settings and show that our
data-driven method can produce superior results without manual intervention, 
at the expense of extensive off-line learning.


\subsection{Deep Learning for Geometric Data}
In the recent years, several methods have been proposed for analyzing and processing
3D shapes by building on the success of machine learning methods, and especially those
based on deep learning (see, for example, recent overviews in
\cite{masci16,bronstein2017geometric}). These methods are based on the notion
that the solutions to many problems in geometric data analysis can benefit from
large data repositories. Many learning-based approaches are aimed at
estimating global properties for tasks such as shape classification and often are
based either on 2D projections (views) of 3D shapes or on volumetric
representations (see, e.g., \cite{qi2016volumetric} for a comparison). However,
several methods have also been proposed for shape segmentation
\cite{guo20153d,maron17} and even shape correspondence
\cite{masci15,wei2016dense,boscaini2016learning}, among many others. 

Although most deep learning-based techniques for 3D shapes heavily exploit the mesh
structure especially for defining convolution operations, few approaches
have also been proposed to directly operate on point clouds. Perhaps the most
closely related to ours are recent approaches of Boulch et
al. \cite{Boulch:2016} and Qi et al. \cite{pointnet17}. The former propose an architecture for estimating unoriented
normals on point clouds by creating a Hough transform-based representation of
local shape patches and using a convolutional neural network for learning normal directions from a
large ground-truth corpus. While not projection based, this method still relies
on a 2D-based representation for learning and moreover loses orientation
information, which can be crucial in practice. More recently, the PointNet architecture~\cite{pointnet17} has been designed to operate on 3D point clouds
directly. This approach is very versatile as it allows to estimate
properties of 3D shapes without relying on 2D projections or volumetric
approximations. The original architecture is purely global, taking the entire
point cloud into account, primarily targeting shape
classification and semantic labeling, and has since then been extended to a
hierarchical approach in a very recent PointNet++ \cite{qi2017}, which is
designed to better capture local structures in point clouds.

Our approach is based on the original PointNet architecture, but rather than using it for
estimating global shape properties for shape classification or semantic
labeling, as has also been the focus of PointNet++ \cite{qi2017}, we adapt it
explicitly for estimating local differential properties including oriented
normals and curvature. 
\section{Overview}
\label{sec:overview}

Estimating local surface properties, such as normals or curvature, from noisy point clouds is a difficult problem that is traditionally solved by extracting these properties from smooth surfaces fitted to local patches of the point cloud. However these methods are sensitive to parameter settings such as the neighborhood size, or the degree of the fitted surface, that need to be carefully set in practice. 
Instead, we propose an alternative approach that is robust to a wide range of conditions with the same parameter settings, based on a deep neural network trained on a relatively small set of shapes.

\begin{figure*}[t!]
	\includegraphics[width=1.0\textwidth]{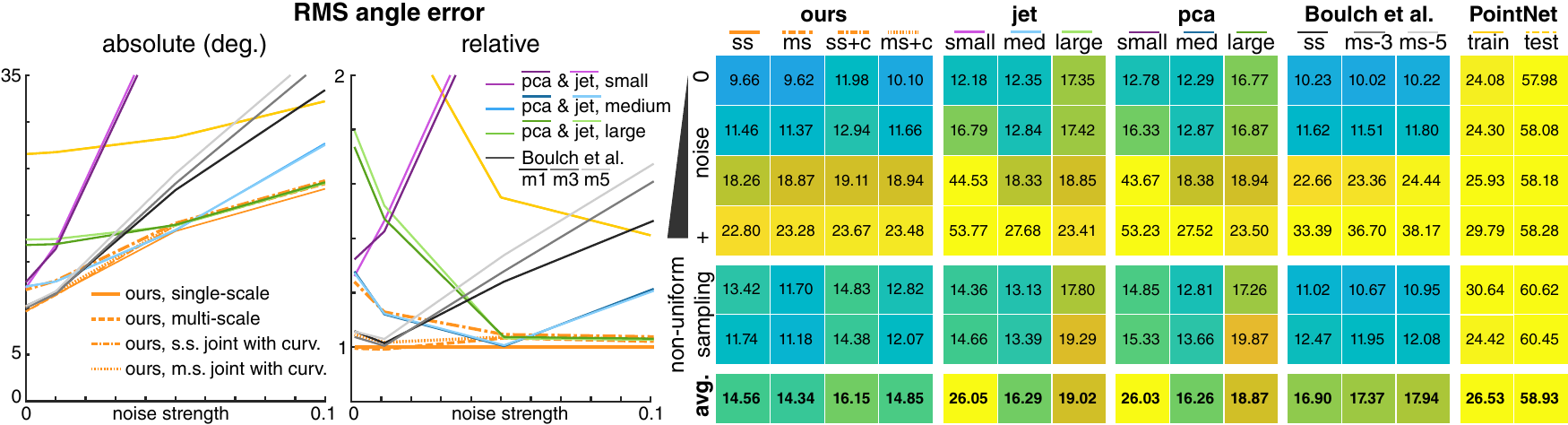}
	\caption{\footnotesize Comparison of the RMS normal angle error of our method (ss: single scale, ms: mult-scale, +c: joint normals and curvature) to geometric methods (jet fitting and PCA) with three patch sizes and two deep learning methods (Boulch et al.~\cite{Boulch:2016} and PointNet~\cite{pointnet17}).
	Note that geometric methods require correct parameter settings, such as the patch size, to achieve good results. Our method gives good results without the need to adjust parameters.}
	\label{fig:comp_unoriented}
\end{figure*}

Given a point cloud $\mathbb{P} = \{p_1, \dots, p_N\}$, our \pcpnet\ network (see Figure~\ref{fig:arch} for an overview) is applied to local patches of this point cloud $\mathbb{P}^r_i \in \mathbb{P}$, centered at points $p_i$ with a fixed radius $r$ proportional to the point cloud's bounding box extent. We then estimate local shape properties at the central points of these patches. The architecture of our network is inspired by the recent PointNet~\cite{pointnet17}, adapted to local $r$-neighborhood patches instead of the entire point cloud. The network learns a set of $k$ non-linear functions in the local patch neighborhoods, which can intuitively be understood as a set of density estimators for different regions of the patch. These give a $k$-dimensional feature vector per patch that can then be used to regress various local features. 

\section{Algorithm}
\label{sec:algorithm}

Our goal in this work is to reconstruct local surface properties from a point cloud that approximately samples an unknown surface. In real-world settings, these point clouds typically originate from scans or stereo reconstructions and suffer from a wide range of deteriorating conditions, such as noise and varying sampling densities.
Traditional geometric approaches for recovering surface properties usually perform reasonably well with the correct parameter settings, but finding these settings is a data-dependent and often difficult task. The success of deep-learning methods, on the other hand, is in part due to the fact that they can be made robust to a wide range of conditions with a single hyper-parameter setting, seemingly a natural fit to this problem. The current lack of deep learning solutions may be due to the difficulty of applying deep networks to unstructured input like point clouds. Simply inputting points as a list would make the subsequent computations dependent on the input ordering.

A possible solution to this problem was recently proposed under the name of PointNet by Qi et al.~\cite{pointnet17}, who propose combining input points with a symmetric operation to achieve order-independence. However, PointNet is applied globally to the entire point cloud, and while the authors do demonstrate estimation of local surface properties as well, these results suffer from the global nature of the method.
PointNet computes two types of features in a point cloud: a single global feature vector for the entire point cloud and a set of local features vectors for each point. The local feature vectors are based on the position of a single point only, and do not include any neighborhood information. This reliance on only either fully local or fully global feature vectors makes it hard to estimate properties that depend on local neighborhood information.

Instead, we propose computing local feature vectors that describe the local neighborhood around a point. These features are better suited to estimate local surface properties.
In this section, we provide an alternative analysis of the PointNet architecture and show a variation of the method that can be applied to local patches instead of globally on the entire point cloud to get a strong performance increase for local surface property estimation, outperforming the current state-of-the art.

\subsection{Pre-processing}
Given a point cloud $\mathbb{P} = \{p_1, \dots, p_N\}$, a local patch $\mathbb{P}^r_i$ is centered at point $p_i$ and contains all points within distance $r$ from the center. Our target for this patch are local surface properties such as normal $n_i$ and principal curvature values $\kappa_i^1$ and $\kappa_i^2$ at the center point $p_i$. To remove unnecessary degrees of freedom from the input space, we translate the patch to the origin and normalize its radius multiplying with $1/r$. Since the curvature values depend on scale, we transform output curvatures to the original scale of the point cloud by multiplying with $r$. Our network takes a fixed number of points as input. Patches that have too few points are padded with zeros (the patch center), and we pick a random subset for patches with too many points.

\begin{figure*}[t!]
	\includegraphics[width=1.0\textwidth]{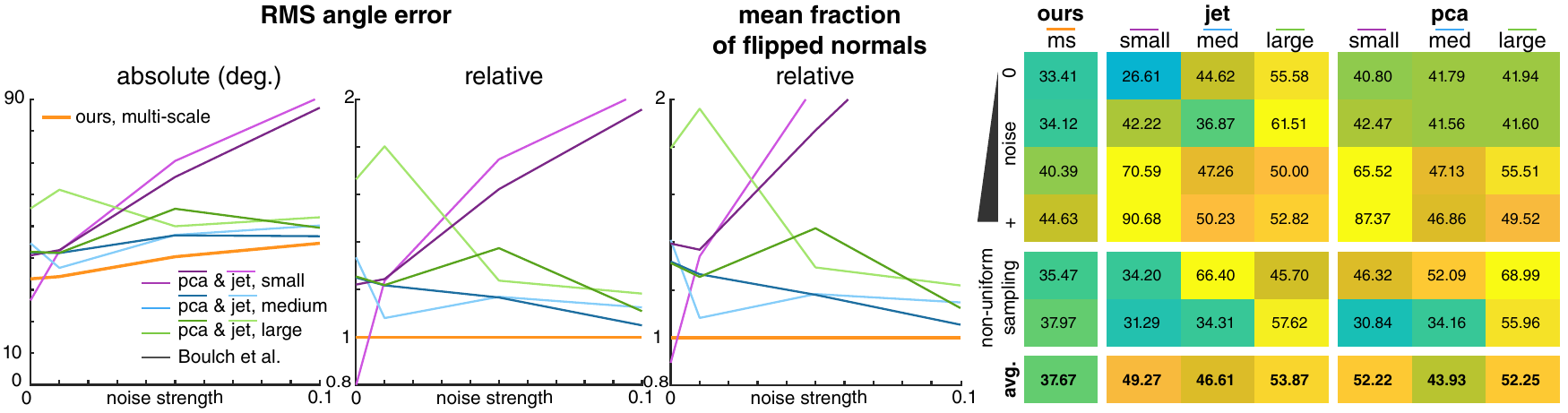}
	\caption{\footnotesize Comparison of our approach for oriented normal estimation with the baseline jet-fitting and PCA coupled with MST-based normal orientation propagation in a post-processing step. We show the RMS angle error, the relative error compared to our performance, as well as relative fraction of flipped normals in other methods across different levels of noise. Note that the errors in oriented normal estimation are highly correlated with the number of orientation flips, suggesting the post-processing step as the main source of error.}
	\label{fig:comp_oriented}
\end{figure*}

\subsection{Architecture}
Our single-scale network follows the PointNet architecture, with two changes: we constrain the first spatial transformer to the domain of rotations and we exchange the $\max$ symmetric operation with a sum. An overview of the architecture is shown in Figure~\ref{fig:arch}. We will now provide intuition for this choice of architecture.

\mypara{Quaternion spatial transformer}
The first step of the architecture transforms the input points to a canonical pose. This transformation is optimized for during training, using a spatial transformer network~\cite{jadberg:2015:st}. We constrain this transformation to rotations by outputting a quaternion instead of a $3 \times 3$ matrix. This is done for two reasons: First, unlike in semantic classification, our outputs are geometric properties that are sensitive to transformations of the patch. This caused unstable convergence behavior in early experiments where scaling would take on extreme values. Rotation-only constraints stabilized convergence. Second, we have to apply the inverse of this transformation to the final surface properties and computing the inverse of a rotation is trivial.

\mypara{Point functions and symmetric operation}
One important property of the network is that it should be invariant to the input point ordering. Qi et al.~\cite{pointnet17} show that this can be achieved by applying a set of functions $\{h_1, \dots, h_k\}$ with shared parameters to each point \emph{separately} and then combine the resulting values for each point using a symmetric operation:
\begin{equation*}
    H_l(\mathbb{P}_i^r) = \sum_{p_j \in \mathbb{P}_i^r} h_l(p_j).
\end{equation*}
$H_l(\mathbb{P}_i^r)$ is then a feature of the patch and $h_l$ are called \emph{point functions}; they are scalar-valued functions, defined in the in the local coordinate frame of the patch. The functions $H_l$ can intuitively be understood as density estimators for a region given by $h_l$. Their response is stronger the more points are in the non-zero region of $h_l$. Compared to using the maximum as symmetric operation, as proposed by Qi et al., our sum did not have a significant performance difference; however we decided to use the sum to give our point functions a simple conceptual interpretation as density estimators.
The point functions $h_l$ are computed as:
\begin{equation*}
    h_l(p_j) = (\mathrm{FNN}_2 \circ \mathrm{STN}_2)\left(g_1(p_j), \dots, g_{64}(p_j)\right),
\end{equation*}
where $\mathrm{FNN}_2$ is a three-layer fully-connected network and $\mathrm{STN}_2$ is the second spatial transformer. The functions $g$ can be understood as a less complex set of point functions, since they are at a shallower depth in the network. They are computed analogous to $h$.


\mypara{Second spatial transformer}
The second spatial transformer operates on the feature vector $\mathbf{g}_j = \left(g_1(p_j), \dots, g_{64}(p_j)\right)$, giving a $64 \times 64$ transformation matrix. Some insight can be gained by interpreting the transformation as a fully connected layer with weights that are computed from the feature vectors $\mathbf{g}$ of \emph{all} points in the patch. This introduces global information into the point functions, increasing the performance of the network.

\mypara{Output regression}
In a trained model, the patch feature vector $\mathbf{H}_j = \left(H_1(\mathbb{P}_i^r), \dots, H_k(\mathbb{P}_i^r)\right)$ provides a rich description of the patch. Various surface properties can be regressed from this feature vector. We use a three-layer fully connected network to perform this regression.

\subsection{Multi-scale}
We will show in the results, that the architecture presented above is very robust to changes in noise strength and sample density. For additional robustness, we experimented with a multi-scale version of the architecture. Instead of using a single patch as input, we input three patches $\mathbb{P}_i^{r1}$, $\mathbb{P}_I^{r2}$ and $\mathbb{P}_I^{r1}$ with different radii. Due to the scale normalization of our patches, these are scaled to the same size, but contain differently sized regions of the point cloud. This allows all point functions to focus on the same region. We also triple the number of point functions, but apply each function to all three patches. The sum however, is computed over each patch separately. This results in nine-fold increase in patch features $H$, which are then used to regress the output properties. Figure~\ref{fig:arch} illustrates a simple version of this architecture with two patches.


\section{Evaluation and Discussion}
\label{sec:results}

In this section, we evaluate our method in various settings and compare it to current state of the art methods. In particular, we compare our curvature estimation to the osculating jets method~\cite{Cazals:2003:EDQ} and a baseline PCA method, and the normal estimation additionally to PointNet~\cite{pointnet17} and the normal estimation method of Boulch et al.~\cite{Boulch:2016}.
We test the performance of these methods on shapes with various noise levels and sampling rates, as described below.
For the method of Boulch et al., we use the code and trained networks provided by the authors. PointNet has code, but no pre-trained network for normal estimation available, so we re-train on our dataset. For the other methods we use the implementation within CGAL~\cite{CGAL}.

By training for different output values, PCPNet can be trained to output 3D normals, curvatures, or both at the same time. We evaluate the performance of the network when trained separately for normals and curvatures or jointly using a single network for both outputs. We also compare the performance of a network trained with both curvature values against networks trained for individual curvature values.

\begin{figure}[t!]
	\includegraphics[width=1.0\columnwidth]{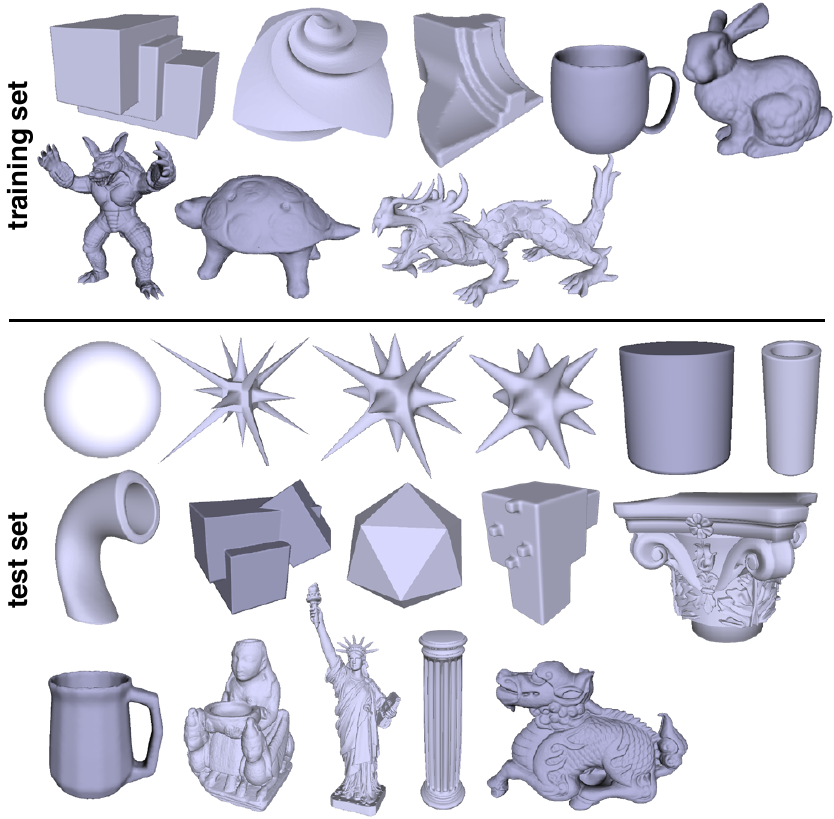}
	\caption{\footnotesize Our shape dataset. We train and test on a mix of simple shapes like the sphere or the boxes, and more detailed shapes like statues or figures. The test set additionally contains three analytic shapes not shown here (sphere, cylinder and a thin sheet shown in Figure~\ref{fig:poisson}).
	}
	\label{fig:dataset}
\end{figure}

\subsection{Dataset}

One of the advantages of training the network on point cloud patches rather than point clouds of complete shapes is that we can sample a large number of patches from each shape. Two patches with near-by center points may still be significantly different from each other in terms of the type of structure they contain, such as edges, corners, etc. Thus, we can use a relatively small dataset of labeled shapes to train our network effectively. Figure \ref{fig:dataset} shows the shapes in our dataset.

Our training dataset contains 8 shapes, half of which are man made objects or geometric constructs with flat faces and sharp corners, and the other half are scans of figurines (bunny, armadillo, dragon and turtle). All shapes are given as triangular meshes.
We sample the faces of each mesh uniformly with $100000$ points to generate a point cloud. Each of these $100000$ points can be used as a patch center. We label each point with the normal of the face it was sampled from. We also estimate the $\kappa^1$ and $\kappa^2$ curvature values at the vertices and interpolate them for each sampled points.
The curvature estimation is performed using the method suggested by Rusinkiewicz in~\cite{Rusinkiewicz:2004:ECA} (the code was provided by authors of~\cite{shabat2015design}).

\begin{figure}[t!]
	\includegraphics[width=1.0\columnwidth]{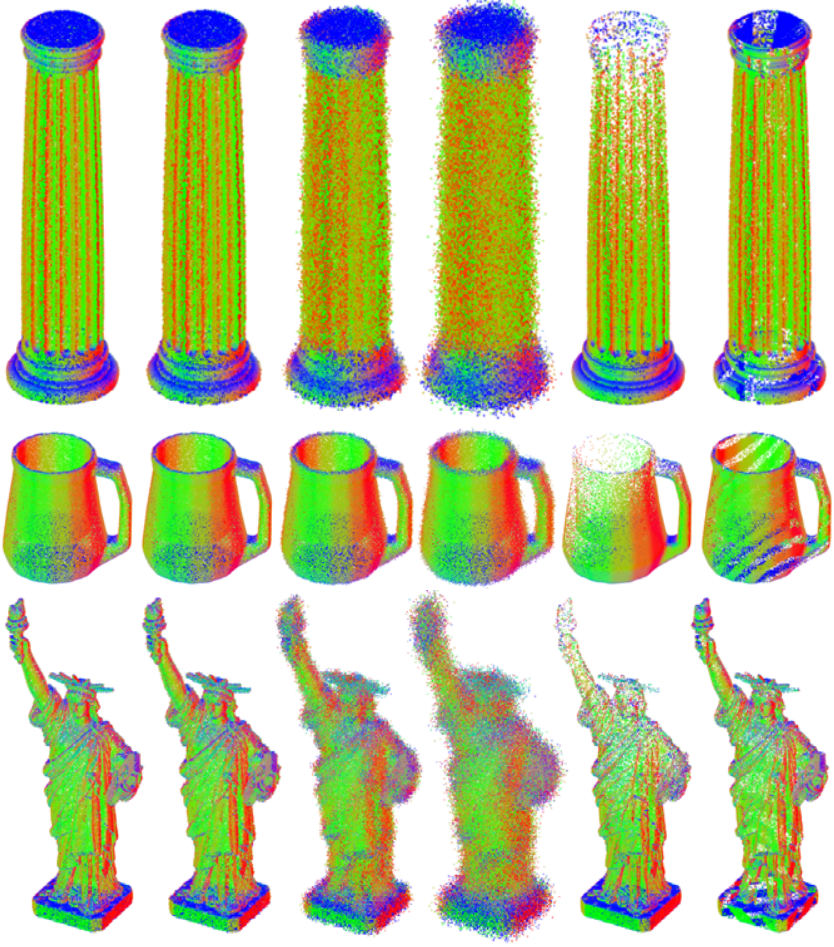}
	\caption{\footnotesize A few point clouds from our dataset with their variants. The base shape is shown in the left column, followed by variants with three noise levels ($0.0025$, $0.012$ and $0.024$), and two non-uniform sampling schemes. Shapes are colored according to their unoriented normals.}
	\label{fig:dataset_detailed}
\end{figure}

For each mesh we generate a noise-free point cloud and three point clouds with added gaussian with a standard deviation of $0.0025$, $0.012$, and $0.024$ of the length of the bounding box diagonal of the shape. Examples are shown in Figure~\ref{fig:dataset_detailed}. The noise is added to the point position but no change is made to ground truth normals and curvatures, as the goal is to recover the original normals and curvatures of the mesh.
In total, our training dataset contains 4 variants of each mesh, or 32 point clouds in total.

Our test set contains 19 shapes with a mix of figurines and man-made objects. We also include 3 shapes that are constructed and sampled analytically from differentiable surfaces which have well-defined normals and curvatures everywhere. For these shapes, the normals and curvatures are computed in an exact manner for each sampled point, rather than being approximated by the faces and vertices of a mesh.

In addition to the three noise variants we described above, we generate two point clouds for each mesh that are sampled with varying density, such that certain regions of the shape are sparsely sampled compared to other regions (see Figure~\ref{fig:dataset_detailed}). This gives us a total of 90 point clouds in the test set.

\begin{figure*}[t!]
	\includegraphics[width=1.0\textwidth]{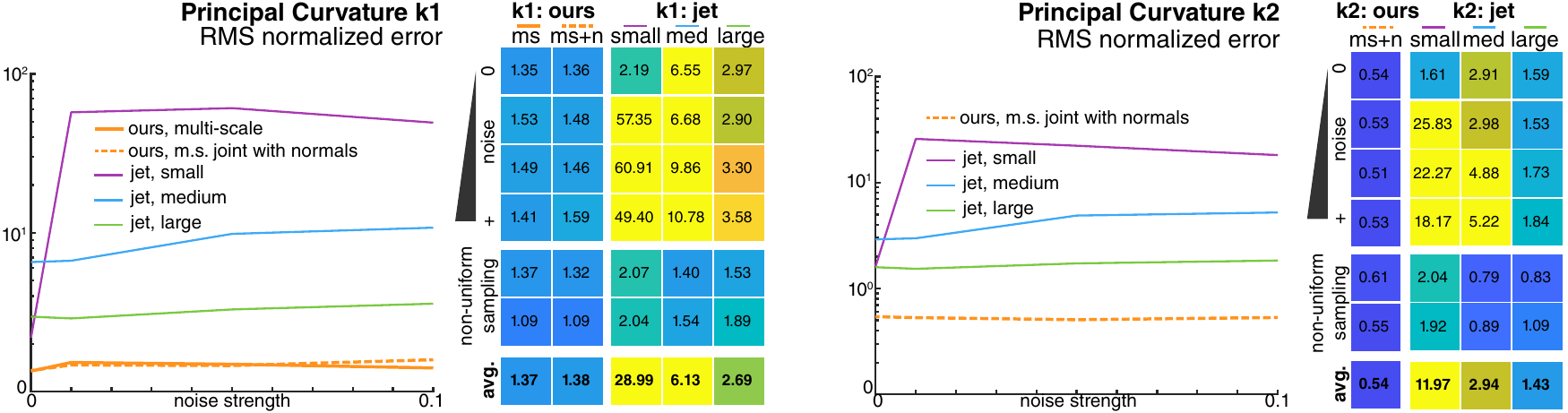}
	\caption{\footnotesize Estimation of both principal curvature values compared to jet fitting. The low performance of jet fitting shows that curvature estimation on point clouds is a challenging problem. Our method achieves significantly better performance, often more than one order of magnitude. Please see Section~\ref{sec:error_metric} for a description of the error metric.\vspace{-3mm}}
	\label{fig:comp_curv}
\end{figure*}

\subsection{Evaluation Metrics}
\label{sec:error_metric}
As loss function during training, and to evaluate our results, we compute the deviation of the predicted normals and/or curvatures from the ground truth data of the shape. 
For normals, we experimented with both the Euclidean distance and the angle difference between the estimated normal and ground truth normal. The mean square Euclidean distance over a batch gave slightly better performance during training, but for better interpretability, we use the RMS angle difference to evaluate our results on the test set.

For curvatures, we compute the mean-square of the following rectified error for training and the RMS for evaluation. The rectified error $L(\Tilde{\kappa})$ is defined as: 
$$ L(\Tilde{\kappa}) = \left|\frac{\Tilde{\kappa}-\kappa}{\max(|\kappa|, 1)}\right|, $$
where $\tilde{\kappa}$ is the estimated curvature and $\kappa$ is the ground truth curvature.
This error is relative to the magnitude of the ground truth curvature, since errors around high curvature areas are expected to be larger.

\subsection{Training and Evaluation Setup}
Our network can be trained separately for normals and curvatures, or jointly to output both at the same time. In the joint network, the loss function is combined during training. We experiment with both variants to test whether the information about the curvatures can help normal estimation and vice versa, since the two are related. We train single-scale and multi-scale networks for each variant, each with $1024$ point functions $h$.

The variants of our network are trained by selecting patches centered randomly at one of the $100$K points in each point cloud. The radius of a patch is relative to the length of the bounding box of the point cloud. The single-scale networks are trained with a patch size of $0.05$, and the multi-scale networks are trained with patch sizes of $0.01$, $0.03$, and $0.07$. We use a fixed number of $500$ points per patch. If there are fewer point within the patch radius, we pad with zeros (the patch center). The network can easily learn to ignore these padded points. If there are more points within the radius, we select a random subset. A more sophisticated subsampling method can be implemented in future work, which may be particularly beneficial for handling varying sampling densities.

In each epoch, we iterate through $1000$ patches from each point cloud in the training set. We train for up to $2000$ epochs on our dataset, or until convergence. Typically training converged before reaching this threshold, taking between $8$ hours for single-scale architectures and $30$ hours for multi-scale architectures on a single Titan X Pascal. A full randomization of the dataset, mixing patches of different shapes in each batch, was vital to achieve stable convergence. All our training was performed in PyTorch~\cite{pytorch} using stochastic gradient descent with batch size $64$, a learning rate of $10^{-4}$ and $0.9$ momentum.

For evaluation, we select a random subset of $5000$ points from each shape in the test set, and output the error of our method and the baseline methods over this subset.
For the CGAL baseline methods~\cite{CGAL}, we use different patch sizes, where the size is determined by the number of nearest neighbors in the patch. For the small patch size we use the recommended setting of 18 nearest neighbors, and we increase the number of nearest neighbors by the same ratio as the area covered by our patches. This amounts to 112 and 450 nearest neighbors for the medium and large patches, respectively.
For Boulch et al.~\cite{Boulch:2016}, we use the single-scale, 3-scale and 5-scale networks provided by the authors.

\begin{figure*}[t!]
	\includegraphics[width=1.0\textwidth]{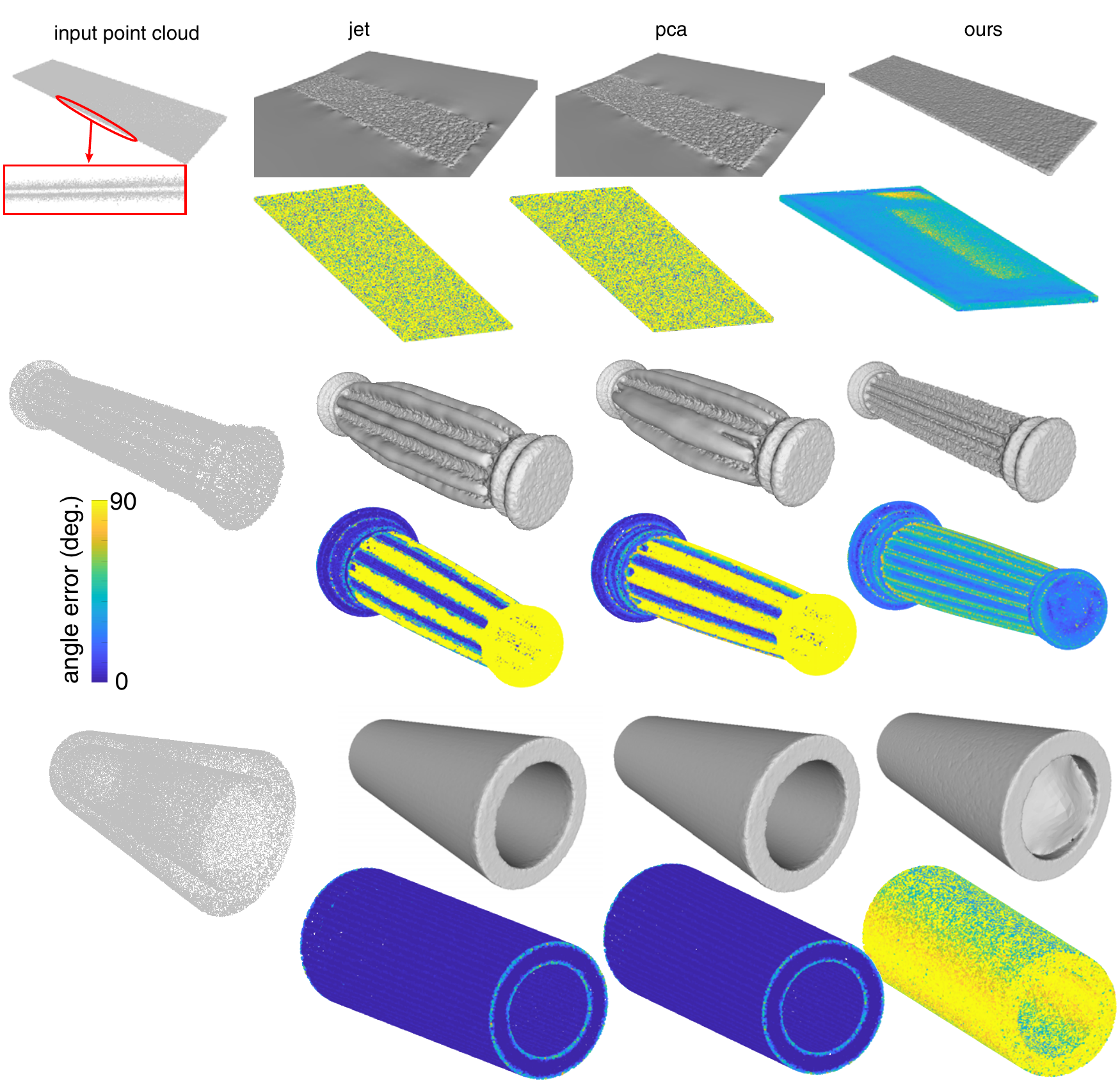}
	\caption{\footnotesize Poisson reconstruction from oriented normals estimated by \pcpnet\ compared to baseline methods. Our method can reconstruct oriented normals with sufficient quality to successfully perform Poisson reconstruction, even in cases that are hard to handle by traditional methods. Although our method performs better on average, there are also failure cases, as demonstrated in the bottom row.
	\vspace{-3mm}}
	\label{fig:poisson}
\end{figure*}

\subsection{Results}

Figure~\ref{fig:comp_unoriented} shows the comparison of unoriented normal estimation using the methods discussed above. In this experiment, we consider either the output normal or the flipped normal, whichever has the lowest error.
In the top section of the table, we show the results for varying levels of noise, from zero noise to high noise. The two rows in the middle show the results for point clouds with non-uniform sampling rate. In each of the categories we show the average for all 20 shapes in the category. The last row shows the global average error over all shapes.
On the left, we show the level of error of each method in relation to the noise level of the input. The graph on the right shows the error of each method relative to the error of our single scale method (marked \emph{ss} in the table).

We can observe the following general trends in these results: first, note that all of our methods consistently outperform competing techniques across all noise levels. In particular, observe that the methods based on jet fitting perform well under a specific intensity of noise, depending on the size of the neighborhood. For example, while jet small works well under small noise, it produces large errors strong noise levels. Conversely, jet large is fairly stable under strong noise but over-smooths for small noise levels, which leads to a large error for clean point clouds. One source of error of our current network compared, e.g. to the results of Boulch et al.~\cite{Boulch:2016} is that our method does not perform as well in the case of changes in sampling density. This is because our network was trained only on uniformly sampled point sets and therefore is not as robust to such changes. Nevertheless, it achieves  comparable results across most levels of non-uniform sampling and shows significant improvement overall. For PointNet, we provide the performance on the training set in addition to the test set. PointNet overfits our training set, and the training set performance gives and approximate lower bound for the achievable error. Since PointNet's point functions cover the whole point cloud, there is less resolution available for local details. This results a large performance gap to other methods on detailed shapes, a gap that gets smaller with stronger noise, where the other methods start to miss details as well.
Note also that our multi-scale architecture produces slightly better results than the single scale one for normal estimation. At the same time, the multi-scale architecture which combines both normals and curvature estimation (ms + c) produces slightly inferior results but has the advantage of outputting both normals and curvature at the same time.

\begin{figure*}[th]
	\includegraphics[width=1.0\textwidth]{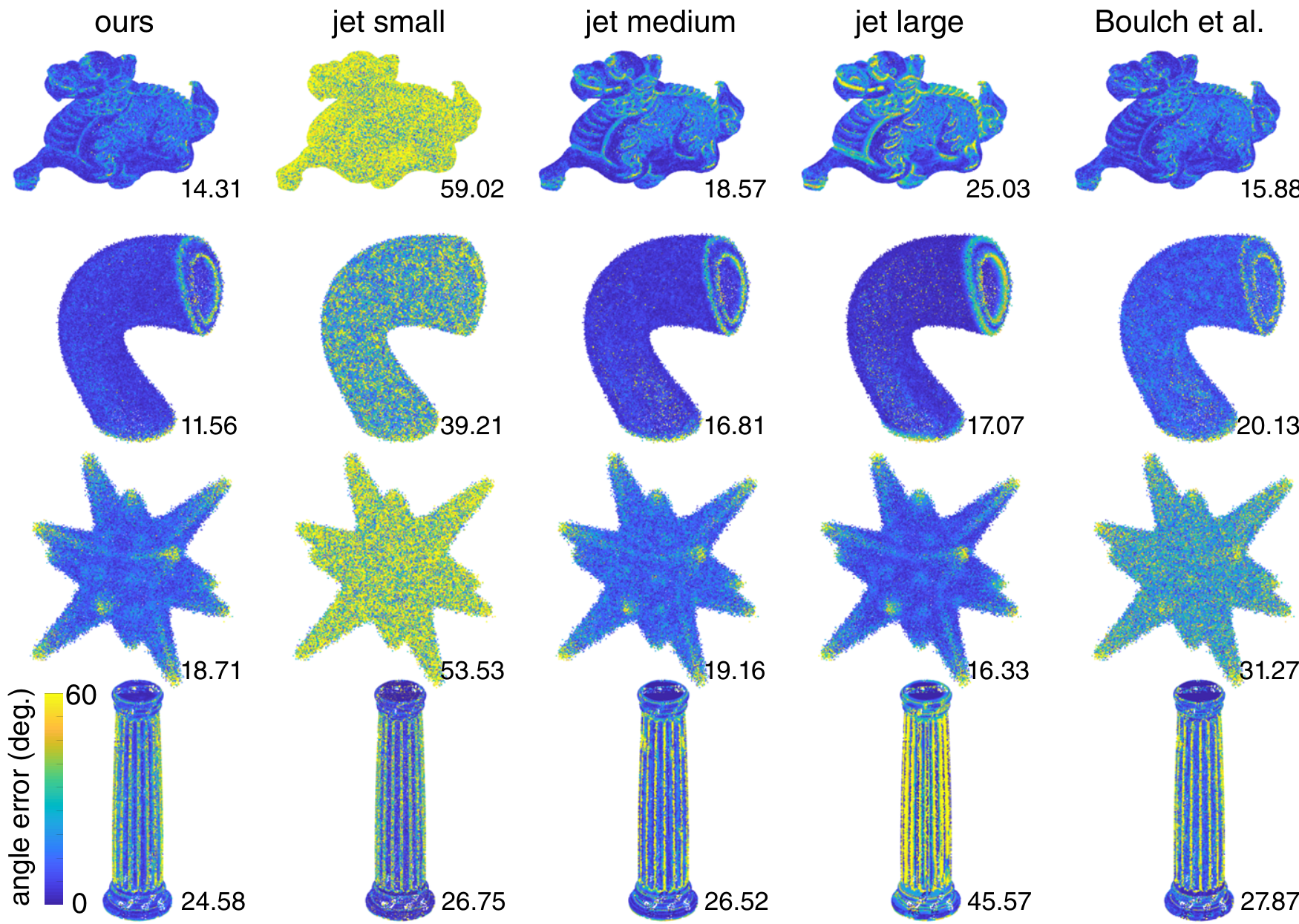}
	\caption{\footnotesize Qualitative comparison of the normals estimated by our method to the normals estimated by baseline methods. Shapes are colored according to normal angle error, where blue is no error and yellow is high error, numbers show the RMS error of the shape.\vspace{-3mm}}
	\label{fig:qual_normals}
\end{figure*}

In Figure \ref{fig:comp_oriented} we show the results of the evaluation of our approach compared to baselines for \emph{oriented} normal estimation. Namely, we use our pipeline directly for estimating oriented normals, and compare them to the jet fitting and PCA together with MST-based orientation propagation as implemented in CGAL~\cite{CGAL}. We do not include the method of Boulch et al.~\cite{Boulch:2016} in this comparison, as it is not designed to produce oriented normals. In Figure \ref{fig:comp_oriented} we plot the RMS angle error while penalizing changes in orientation as well as the error relative to our performance. Finally, we also report the relative fraction of normals that are flipped with respect to the ground truth orientation, for other methods across different levels of noise. Note that the errors in oriented normal estimation are highly correlated with the number of orientation flips, suggesting that the orientation propagation done during the post-processing is the main source of error. Also note that orientation propagation only works well for extremely small noise levels and very quickly deteriorates leading to large errors. Given a set of oriented normals, we can also perform Poisson reconstruction. Figure~\ref{fig:poisson} shows a few examples of objects reconstructed with our oriented normals. Note that the top-most point cloud samples two sides of a thin sheet, making it hard to determine which side of the sheet a point originated from. In the middle row, the point cloud exhibits sharp edges that are hard to handle for the MST-based propagation. The bottom-most shape shows a failure case of our method. We suspect that a configuration like the inside of this pipe was not encountered during training. Increasing or diversifying the training set should solve this problem. Note that
on average, we expect the performance of our method for Poisson reconstruction to be proportional to the fraction of flipped normals, as shown in Figure~\ref{fig:comp_oriented}.

Figure~\ref{fig:comp_curv} shows the comparison of our curvature estimation to jet fitting. Due to the sensitivity of curvature to noise, noisy point clouds are a challenging domain for curvature estimation. Even though our method does not achieve a very high level of accuracy (recall that the error is normalized by the magnitude of the ground truth curvature), our performance is consistently superior to jet fitting on both principal curvature values, by a significant margin.

\begin{figure*}[th]
	\includegraphics[width=1.0\textwidth]{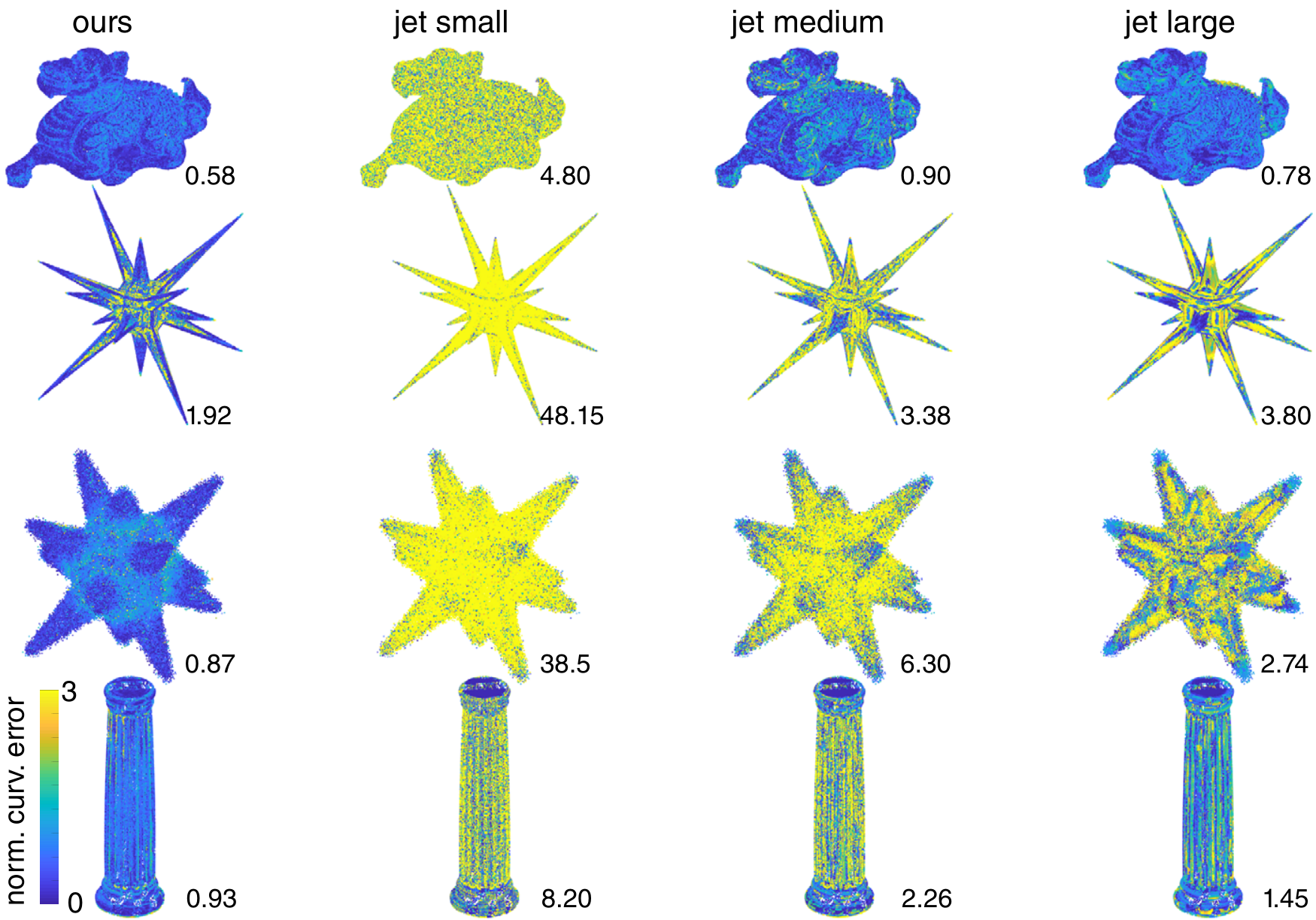}
	\caption{\footnotesize Qualitative comparison of curvature values estimated by our method to the curvature values estimated by a baseline method with three different patch sizes. Shapes are colored according to the curvature estimation error, where blue is no error and yellow is high error, numbers show the RMS error of the shape.}
	\label{fig:qual_curv}
\end{figure*}

Qualitative comparisons of the normal error on four shapes of our dataset are shown in Figure~\ref{fig:qual_normals}. Note that for classical surface fitting (jet in these examples), small patch sizes work well on detailed structures like in the bottom row, but fail on noisy point clouds, while large patches are more tolerant to noise, but smooth out surface detail. Boulch et al.~\cite{Boulch:2016} perform well in a low-noise setting, but the performance quickly degrades with stronger noise. Even though our method does not always perform best in cases that are optimal for the parameter setting of a another method (e.g. versus jet fitting with large patch size in the third row, where the shape is very smooth and noisy), it performs better in most cases and on average.

In Figure~\ref{fig:qual_curv}, we show qualitative results of curvature estimation for selected shapes. The color of the points marks the error in curvature, where blue is no error and yellow is high error. Errors are computed in the same manner as described in Section~\ref{sec:error_metric}, and are clamped between $0$ and $3$. The error of our curvature estimation is typically below $1$, while for previous method the estimation is often orders of magnitude higher than the ground truth curvature.



\section{Conclusion, Limitations, and Future Work}
\label{sec:conclusion}

We presented a unified method for estimating oriented normals and principal curvature values in noisy point clouds. Our approach is based on a modification of a recently proposed PointNet architecture, in which we place special emphasis on extracting local properties of a patch around a given central point. We train the network on point clouds arising from triangle meshes corrupted by various levels of noise and show through extensive evaluation that our approach achieves state-of-the-art results across a wide variety of challenging scenarios. In particular, our method allows to replace the difficult and error-prone manual tuning of parameters, present in the majority of existing techniques with a data-driven training. Moreover, we show improvement with respect to other recently proposed learning-based methods.

While producing promising results in a variety of challenging scenarios, our method can still fail in some settings, such as in the presence of large flat areas, in which patch-based information is not enough to determine the normal orientation. For example, our oriented normal estimation procedure can produce inconsistent results, e.g., in the centers of faces a large cube. A more in-depth analysis and a better-adapted multi-resolution scheme might be necessary to alleviate such issues.

In the future, we also plan to extend our technique to estimate other differential quantities such as principal curvature directions or even the full first and second fundamental forms, as well as other mid-level features such as the Shape Diameter Function from a noisy incomplete point cloud. Finally, it would also be interesting to study the relation of our approach to graph-based neural networks~\cite{
defferrard2016convolutional,henaff2015deep} on graphs built from local neighborhoods of the point cloud.

\section*{Acknowledgement} This work was supported by the ERC Starting Grants SmartGeometry (StG-2013-335373) and EXPROTEA (StG-2017-758800), the Open3D Project (EPSRC Grant EP/M013685/1), the Chateaubriand Fellowship, chaire Jean Marjoulet from Ecole Polytechnique, FUI project TANDEM 2, and a Google Focused Research Award.


\bibliographystyle{eg-alpha-doi}
\bibliography{pcpnet}

\end{document}